\documentclass[twocolumn]{aastex63}
\usepackage{amsmath}
\usepackage{color}
\usepackage{natbib}
\usepackage{enumitem}
\usepackage{float}
\usepackage{hyperref}
\usepackage{listings}
\usepackage{fontawesome}
\usepackage{xcolor}
\usepackage{xspace} 
\usepackage{url}
\usepackage{fontawesome}

\definecolor{linkcolor}{rgb}{0.,0.3,0.7}

\definecolor{codegreen}{rgb}{0,0.6,0}
\definecolor{codegray}{rgb}{0.5,0.5,0.5}
\definecolor{codepurple}{rgb}{0.58,0,0.82}
\definecolor{backcolour}{rgb}{0.95,0.95,0.92}

\lstdefinestyle{mystyle}{
    commentstyle=\color{codegray},
    keywordstyle=\color{codegreen},
    numberstyle=\tiny\color{gray},
    stringstyle=\color{codepurple},
    basicstyle=\ttfamily\footnotesize,
    breakatwhitespace=false,         
    breaklines=true,                 
    captionpos=b,                    
    keepspaces=true,                 
    numbers=none,                    
    numbersep=5pt,                  
    showspaces=false,                
    showstringspaces=false,
    showtabs=false,                  
    tabsize=2
}

\hypersetup{breaklinks, colorlinks,
  urlcolor=linkcolor,linkcolor=linkcolor,citecolor=linkcolor}
\newcommand{\code}[1]{\texttt{#1}}

\newcommand{\codeicon}{{\color{linkcolor}\faFileCodeO}}

\newcommand{\artpop}{\texttt{ArtPop}\xspace}
\newcommand{\sersic}{S\'ersic\xspace}
\newcommand{\script}[1]{\href{https://github.com/ArtificialStellarPopulations/artpop-paper-figures/blob/main/scripts/#1}{{\codeicon}}}
\newcommand{\readthedocs}[1]{\href{https://artpop.readthedocs.io/en/latest/tutorials/#1.html}{{\faBook}}}
\newcommand{\github}[1]{\href{https://github.com/ArtificialStellarPopulations/ArtPop/tree/main/src/artpop/#1}{{\faGithub}}}
\newcommand{\githubmain}{\href{https://github.com/ArtificialStellarPopulations/ArtPop/}{{\faGithub}}}
\newcommand{\home}{\href{https://artpop.readthedocs.io/en/latest/}{\faHome}}

\submitjournal{ApJ}

\graphicspath{{figures/}}

\shorttitle{ArtPop}

\begin{document}\sloppy\sloppypar\raggedbottom\frenchspacing 

\title{ArtPop: A Stellar Population and Image Simulation Python Package}

\author[0000-0003-4970-2874]{Johnny P. Greco}
\altaffiliation{NSF Astronomy \& Astrophysics Postdoctoral Fellow}
\affiliation{Center for Cosmology and AstroParticle Physics (CCAPP), The Ohio State University, Columbus, OH 43210, USA}

\author[0000-0002-1841-2252]{Shany Danieli}
\altaffiliation{NASA Hubble Fellow}
\affiliation{Department of Astrophysical Sciences, 4 Ivy Lane, Princeton University, Princeton, NJ 08544}
\affiliation{Institute for Advanced Study, 1 Einstein Drive, Princeton, NJ 08540, USA}

\begin{abstract}

We present \code{Art}ificial Stellar \code{Pop}ulations (\artpop), an open-source Python package for synthesizing stellar populations and generating artificial images of fully populated stellar systems. The code is designed to be intuitive to use and as modular as possible, making it possible to use each of its functionalities independently or together. \artpop has a wide range of scientific and pedagogical use cases, including the measurement of detection efficiencies in current and future imaging surveys, the calculation of integrated stellar population parameters, quantitative comparisons of isochrone models, and the development and validation of astronomical image processing algorithms. In this paper, we give an overview of the \artpop package, provide simple coding examples to demonstrate its implementation, and present results from some potential applications of the code. We provide links to the source code that created each example and figure throughout the paper. \artpop is under active development, and we welcome bug reports, feature requests, and code contributions from the community. \home\ \githubmain\ \readthedocs{quickstart} \script{}
\end{abstract}


\section{Introduction} \label{sec:intro}

The physical properties of galaxies are almost entirely derived from observations of their light. At distances too large for resolving galaxies into individual stars, stellar population synthesis modeling of the spectral energy distribution may be used to infer integrated properties such as the total stellar mass, star formation rate, and metal content \citep[][and references therein]{Walcher:2011, Conroy:2013}. In the nearby universe, on the other hand, it is possible to study galaxies using observations of individual stars. Star formation histories (SFHs), for instance, can be measured from Hubble Space Telescope (HST)-based resolved star color-magnitude diagrams (CMDs; e.g., \citealt{Weisz:2008}), though such measurements are resource intensive and are only possible for a very small subset of the galaxy population. 

In between these two extremes lies the semi-resolved regime. The statistical “lumpiness” observed in relatively nearby galaxies, which are unresolved due to stellar crowding, contains a plethora of information about the underlying stellar populations. For instance, the sparse number of red giants relative to main sequence stars produces surface brightness fluctuations (SBFs) between pixels, which are well-known as an accurate extragalactic distance indicator \citep{1988AJ.....96..807T}, as well as a sensitive probe of stellar populations \citep[e.g.,][]{Buzzoni:1993, Ajhar:1994, Cantiello:2003, Raimondo:2005}. Taking this idea even further, fluctuation spectroscopy \citep{vD:2014} provides a powerful method for studying metal-rich giants in massive elliptical galaxies, and pixel CMDs \citep{Conroy:2016} generalize the SBF technique and are sensitive to SFHs, distances, and metallicities, as demonstrated in \citet{Cook:2019, Cook:2020}.

Current and future wide-field imaging surveys will revolutionize our view of extragalactic stellar systems by producing large sets of deep and high resolution images. Indeed, surveys like the Hyper Suprime-Cam (HSC) Subaru Strategic Program \citep{2018PASJ...70S...4A} and the Dark Energy Survey \citep[DES;][]{Abbott:2018} are already producing large catalogs of low surface brightness galaxies over large areas of the sky (\citealt{Greco:2018}; \citealt{Tanoglidis:2021}), an effort that will be continued in earnest by the Vera Rubin Observatory Legacy Survey of Space and Time \citep[LSST;][]{Ivezic:2019}. From space, the Nancy Grace Roman Space Telescope \citep{Spergel:2015} will deliver HST-quality imaging  over wide areas, which will significantly increase the number of systems for which it will be possible to study stellar populations in both the resolved and semi-resolved regimes.

Artificial image simulations have long been an invaluable tool in the preparation for and exploitation of unprecedented data sets. To demonstrate how the then-hypothetical HST might supplement ground-based studies, \citet{1988IAUS..126..455B} simulated HST observations of a nearby globular cluster. The same simulation software was later used in \citet{1988AJ.....96..807T} to quantify the feasibility and limitations of the SBF distance measurement method. Today, similar star-by-star image simulations have become the gold standard for measuring point source completeness and modeling resolved stellar populations \cite[e.g.,][]{Stetson:1987, Dolphin:2000, Dalcanton:2009}.

In this work, we present \code{Art}ificial Stellar \code{Pop}ulations (\artpop), an open-source Python package for modeling stellar populations and generating their corresponding images. \artpop was conceptually introduced in \citet{Danieli:2018}, and it soon proved useful in helping to confirm the distance to the ``galaxy lacking dark matter'' \citep{vD-DF2-distance-2018}. \artpop was then generalized, expanded upon, and used in \citet{Greco:2021} to study SBFs in low-luminosity stellar systems.

The goal of \artpop is to provide the community with a user-friendly tool for generating artificial images of resolved and semi-resolved stellar systems. It complements more comprehensive packages like \code{GalSim} \citep{Rowe:2015}, which tend to have steep learning curves and have most often been applied to studies of unresolved galaxies. It is our hope that \artpop will be useful both for scientific applications and in the classroom, where it can provide students with a unique perspective of stellar populations and their image formation.

The paper is organized as follows. In Section \ref{sec:methods}, we present our methods for synthesizing stellar systems and generating artificial images. In Section \ref{sec:code}, we describe the \artpop software package. We include links (indicated by the \codeicon\ icon) to the Python code that created each example and figure. In Section \ref{sec:examples}, we provide a diverse set of scientific and pedagogical example applications. Finally, we conclude with a summary in Section \ref{sec:summary}.

\section{Synthesizing Stellar Systems} \label{sec:methods}

There are three main components that are necessary for generating artificial images of fully-populated stellar systems. The first is modeling stellar populations and the associated stellar fluxes. To synthesize complex star formation histories, multiple single-burst populations may be combined. The second is spatial information, including the distance to the system and image coordinates for all its stars. Finally, image processing tools are required to inject the stellar fluxes into an image, convolve with a point-spread function (PSF), and add noise according to a set of instrumental and observational parameters. In this section, we describe how we have implemented each of these components. 

\subsection{Stellar Populations}\label{sec:pops}

The basic building block of stellar population synthesis is the simple stellar population (SSP), which consists of a population of stars born at the same time with a single metallicity and abundance pattern. To build SSPs, \artpop starts from pre-calculated stellar isochrones and synthetic photometry. While the code can work with any set of models, here we use the  Modules for  Experiments  in  Stellar  Astrophysics \citep[MESA;][]{Paxton:2011, Paxton:2013, Paxton:2015} Isochrones and Stellar  Tracks (MIST) project\footnote{\url{http://waps.cfa.harvard.edu/MIST/}} \citep{Dotter:2016, Choi:2016}. In particular, we use synthetic photometry generated from the rotating models with $v/v_\mathrm{crit}=0.4$ from MIST version 1.2.

Given a set of stellar isochrones, we populate an SSP by sampling stellar masses from the initial mass function, $\Phi(M_i)$, using inverse transform sampling. The form of $\Phi(M_i)$ is an optional parameter in \artpop. Unless noted otherwise in this work, we assume the initial mass function from \citet{Kroupa:2001}, with a minimum mass of $M_\mathrm{min}=0.1~M_\odot$. The maximum mass of sampled stars is set by the stellar isochrone at a given age and metallicity. The total mass of the SSP, including both stars and stellar remnants, is given by
\begin{equation}\label{eqn:total-mass}
    M_\mathrm{tot}(t) = \int_{M_\mathrm{min}}^{M(t)} M_\mathrm{act}(M_i) \Phi(M_i)\,dM_i + M_\mathrm{rem},
\end{equation}
where $t$ is the stellar population age, $M_\mathrm{act}(M_i)$ is the actual mass of a star that had an initial mass $M_i$, and $M_\mathrm{rem}$ is the mass contained in stellar remnants including white dwarfs, neutron stars, and black holes. For the normalization of the initial mass function, we use a maximum mass of $M_\mathrm{max}=120~M_\odot$.

We assign masses to stellar remnants using the prescription of \citet{Renzini:1993}. For stars with initial masses $M_i<8.5~M_\odot$, the remnants are assumed to be white dwarfs of mass $M_\mathrm{rem} = 0.077\,M_i + 0.48$. Stars with $8.5~M_\odot \leq M_i < 40~M_\odot$ leave behind 1.4~$M_\odot$ neutron stars. Finally, the most massive stars with initial masses $M_i \geq 40~M_\odot$ produce black holes of mass 0.5\,$M_i$. These initial-final mass relations are of course rough approximations, but they are standard in the field of stellar population synthesis \citep[e.g.,][]{BC03, Maraston:1998, Conroy:2009}, making it possible to interpret and compare \artpop models in this context. 

In practice, \artpop only samples ``live'' stars, which have masses that are included in the isochrone. To account for stellar remnants in the total mass that is reported by the \artpop model, we apply a correction factor given by the ratio of the mass---as defined in Equation~(\ref{eqn:total-mass})---with and without remnants. This factor generally ranges from total masses (live stars and remnants) that are ${\sim}5\%$-$50\%$ larger than the sum of the sampled stellar masses, depending on the age and metallicity of the SSP. Of course,  older SSPs, in which stars have more time to evolve, have more mass locked up in stellar remnants than their younger counterparts.

\begin{figure*}[t!]  
  \centering
  \includegraphics[width=\textwidth]{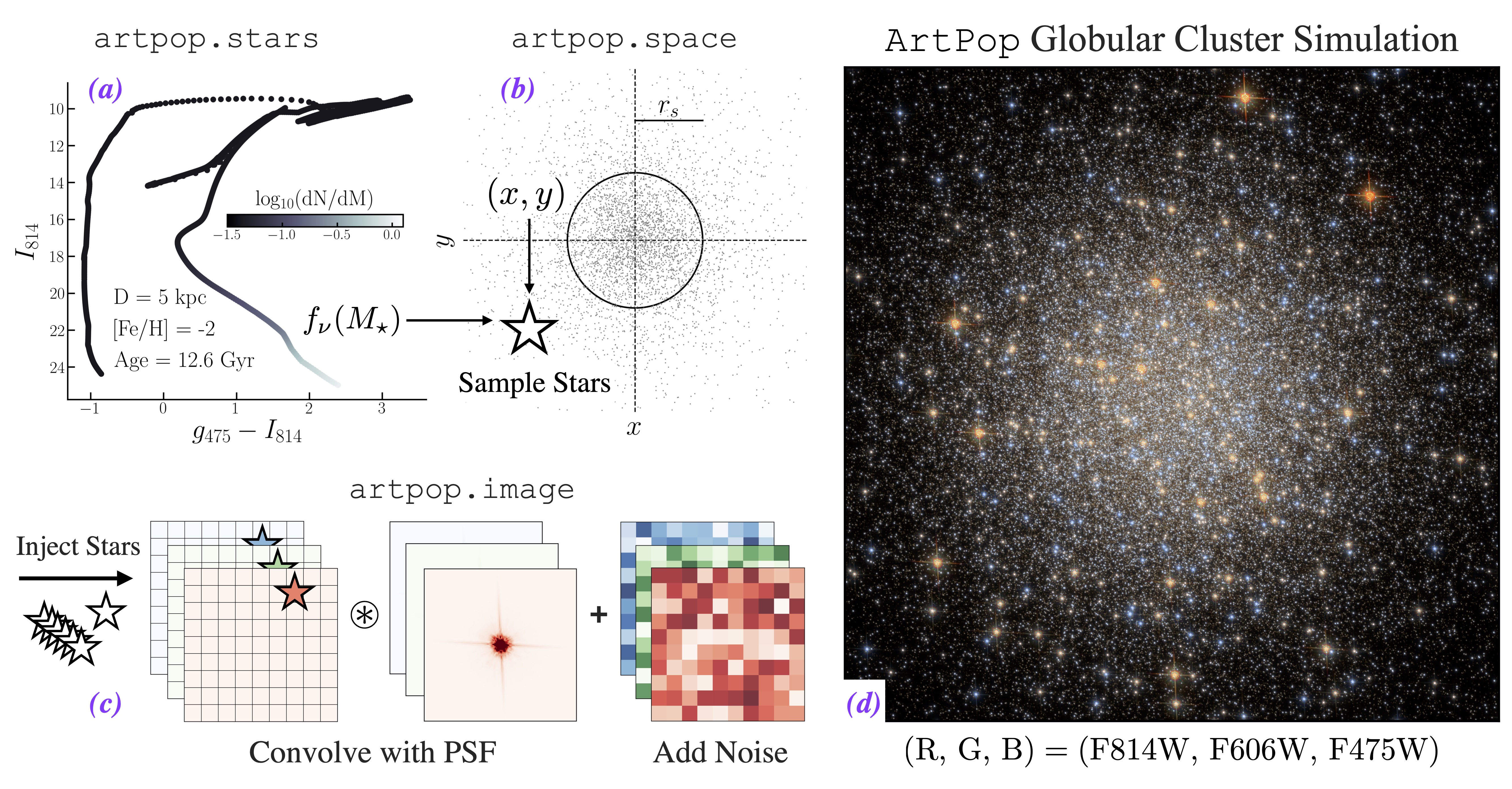}
      \caption{A schematic overview of the procedure we follow to generate artificial images of fully-populated stellar systems. In (a), we sample stellar masses from an initial mass function (weights indicated by the color bar) and interpolate the associated stellar fluxes from an isochrone of the indicated age and metallicity. In (b), we assign spatial positions to each star using random draws from a two-dimensional Plummer distribution. In (c), the stars are injected into an image array, which is then convolved with the point-spread function (PSF). Finally, noise from the sources, sky, and detector are added to the image. In (d), we show a $grI$-composite image of an \artpop simulation of a metal poor globular cluster at 5~kpc. In this example, the mock observations used the filter set, PSFs, and pixel scale of HST ACS/WFC. \script{figure_1_greco_and_danieli.py}}
  \label{fig:diagram}
\end{figure*}

\subsection{Spatial Information}\label{sec:spatial}

In addition to the stellar population synthesis described in the previous section, we must specify the spatial properties of the stars to completely describe the system. In particular, we need to know the stellar density distribution in physical units, the system distance to convert relative positions to angular units and stellar luminosities to brightnesses, and the pixel scale of the mock imaging system to convert the stellar positions to image coordinates. \artpop can inject stars into images using arbitrary user-provided image coordinates, but we provide functions for sampling random positions from common spatial distributions. 

Currently, \artpop has sampling functions for a uniform surface brightness distribution, the Plummer distribution \citep{Plummer:1911}, which provides a good description of the distribution of mass in globular clusters, and the more general S\'{e}rsic distribution \citep{Sersic:1968}, which spans the range of observed concentrations in stellar density distributions. An important note about sampling stellar positions is that the maximum radius out to which stars are sampled should be at least several times the scale length of the spatial distribution, even if such stars will fall outside of the image. Otherwise, too many stars will be sampled near the center of the galaxy, leading to a surface brightness distribution that is inconsistent with the input parameters.

\subsection{Artificial Images}\label{sec:art-images}

Given a simulated stellar population, including the positions and fluxes of every star, \artpop generates an artificial image by injecting the individual stellar fluxes into an image array and convolving with a point-spread function (PSF). The current implementation of \artpop assumes each star falls within the center of a pixel. For most purposes, this simplification will have a negligible effect, provided the PSF is well-sampled \citep[e.g.,][]{Olsen:2003}; to the extent that spatial sampling matters for a given application, its impact will be most acute in rare stellar phases for which the density is such that we expect few stars per pixel. We plan to include optional subpixel star injection in a future version of \artpop.

There are several options for adding noise to an \artpop simulation, spanning the noiseless ``ideal'' case to fully artificial images with read noise and Poisson noise from the sky and artificial sources. In the latter case, the user must provide the necessary instrumental and observational parameters, including the aperture diameter, detector and photometric filter properties, exposure time, and sky surface brightness. 

To convert stellar magnitudes in bandpass $x$ to photon counts, we use the analytic expression 
\begin{equation}\label{eqn:counts}
	C_x = A\cdot\varepsilon_x\cdot\frac{t_\mathrm{exp}}{h}\frac{\Delta \lambda_x}{\lambda_{\mathrm{eff,}\,x}}10^{-0.4\,(m_x\,+\, 48.6)},
\end{equation}
where $A$ is the effective collecting area of the telescope, $\varepsilon_x$ is the efficiency in band $x$ (set to unity by default), $t_\mathrm{exp}$ is the exposure time, $h$ is the Planck constant, $\Delta \lambda_x$ and $\lambda_{\mathrm{eff,}\,x}$ are the width and effective wavelength of bandpass $x$, and $m_x$ is the stellar AB  magnitude \citep{Oke:1983} in bandpass $x$. 

After convolution with the PSF, Poisson noise is generated from the combined counts of the source and sky, and the read noise is assumed to be Gaussian. If the galaxy is injected into a real image---which will already have detector and sky noise---Poisson noise is optionally generated from only the source counts before converting into the image flux units, provided the necessary parameters for Equation~(\ref{eqn:counts}) are given as input.     

\section{The Software Package} \label{sec:code}

In this section, we give an overview of the \artpop software package and provide coding examples to demonstrate the code implementation and basic usage. The code is written in the Python programming language and is entirely open source. We refer the reader to the project website\footnote{\url{https://artpop.readthedocs.io}} for a detailed description of the code, installation instructions, and a growing list of tutorials. We are actively developing \code{ArtPop} in a public GitHub repository\footnote{\url{https://github.com/ArtificialStellarPopulations/ArtPop}} and welcome bug reports, feature requests, and code contributions from the community. Throughout the paper, we provide links (as \codeicon\ icons) to the Python code that created each figure and example.

\subsection{Code Structure} \label{sec:code-structure}

An important feature of the code design is that it is highly modular and extensible. This makes it possible for each of \artpop's functionalities to be used independently, together, or in combination with independently-generated input data (e.g., stellar positions and/or fluxes). \artpop is divided into three primary modules---one for each of the components necessary for generating artificial images of fully-populated stellar systems, as described in Section~\ref{sec:methods}. In particular, the core of the code is composed of the \code{artpop.stars}, \code{artpop.space}, and \code{artpop.image} modules, which are used for building stellar populations, sampling spatial distributions, and generating artificial images, respectively. 

In Figure~\ref{fig:diagram}, we show a schematic overview of the procedure we follow to generate an artificial HST-like image of a globular cluster. For each step, we indicate the \artpop module in which the corresponding code is implemented. In (a), we use the \code{artpop.stars} module to sample stellar masses from a user-specified initial mass function (indicated by the color bar) and interpolate the associated fluxes from a stellar isochrone. In (b), each star is assigned a spatial position in image coordinates using the \code{artpop.space} module. The stellar fluxes and positions are stored in an \code{artpop.Source} object, which in (c), is ``observed'' using the \code{artpop.image} module. Finally, the \artpop image simulation is shown as a $grI$-composite image in (d).

\subsection{Coding Example: Stellar Populations}\label{sec:code-ssp}

A useful example of \artpop's modularity is using the \code{artpop.stars} module to generate simple and composite stellar populations. This capability is independent from making images and may be used to calculate integrated population parameters such as total magnitude, color, and the surviving stellar mass. Given a user-specified initial mass function and isochrone model, \artpop can calculate such parameters either using numerical integration or by sampling a finite number of stars. The latter method introduces stochasticity at low stellar mass, which may be desired in certain applications.

Calculations involving stellar isochrones are carried out in the flexible \code{Isochrone} class. Three input arguments are required to initialize an \code{Isochrone} object: SSP initial stellar masses (\code{mini}), the actual stellar mass after accounting for mass loss (\code{mact}), and a table of the associated stellar magnitudes (\code{mags}). Importantly, the code implementation is independent from how these parameters were generated, provided they are given in the correct format. Assuming these arguments have been defined, the code may be implemented as follows \script{section_3_greco_and_danieli.py}: 
\begin{lstlisting}[language=Python]
from artpop.stars import Isochrone

iso = Isochrone(mini, mact, mags)
\end{lstlisting} 
The \code{iso} object has methods for performing real-time calculations of integrated SSP parameters. For example, if the magnitude table contains LSST magnitudes, the IMF-weighted $g-i$ color and surviving stellar mass (assuming a Salpeter initial mass function) of the SSP may be calculated using
\begin{lstlisting}[language=Python]
g_i =  iso.ssp_color(
    blue="LSST_g", # blue filter
    red="LSST_i",  # red filter
    imf="salpeter" # initial mass function
)
m_survive = iso.ssp_surviving_mass("salpeter")
\end{lstlisting} 
For convenience, we have implemented a helper MIST-specific class, \code{MISTIsochrone}, which inherits all the methods from \code{Isochrone}. The user provides the desired SSP age, metallicity, and photometric system, and \code{MISTIsochrone} loads the required input parameters using the MIST synthetic photometry grids\footnote{\url{http://waps.cfa.harvard.edu/MIST/model_grids.html}}, interpolating over metallicity if necessary.

\begin{figure}[t!]
  \centering
  \includegraphics[width=\columnwidth]{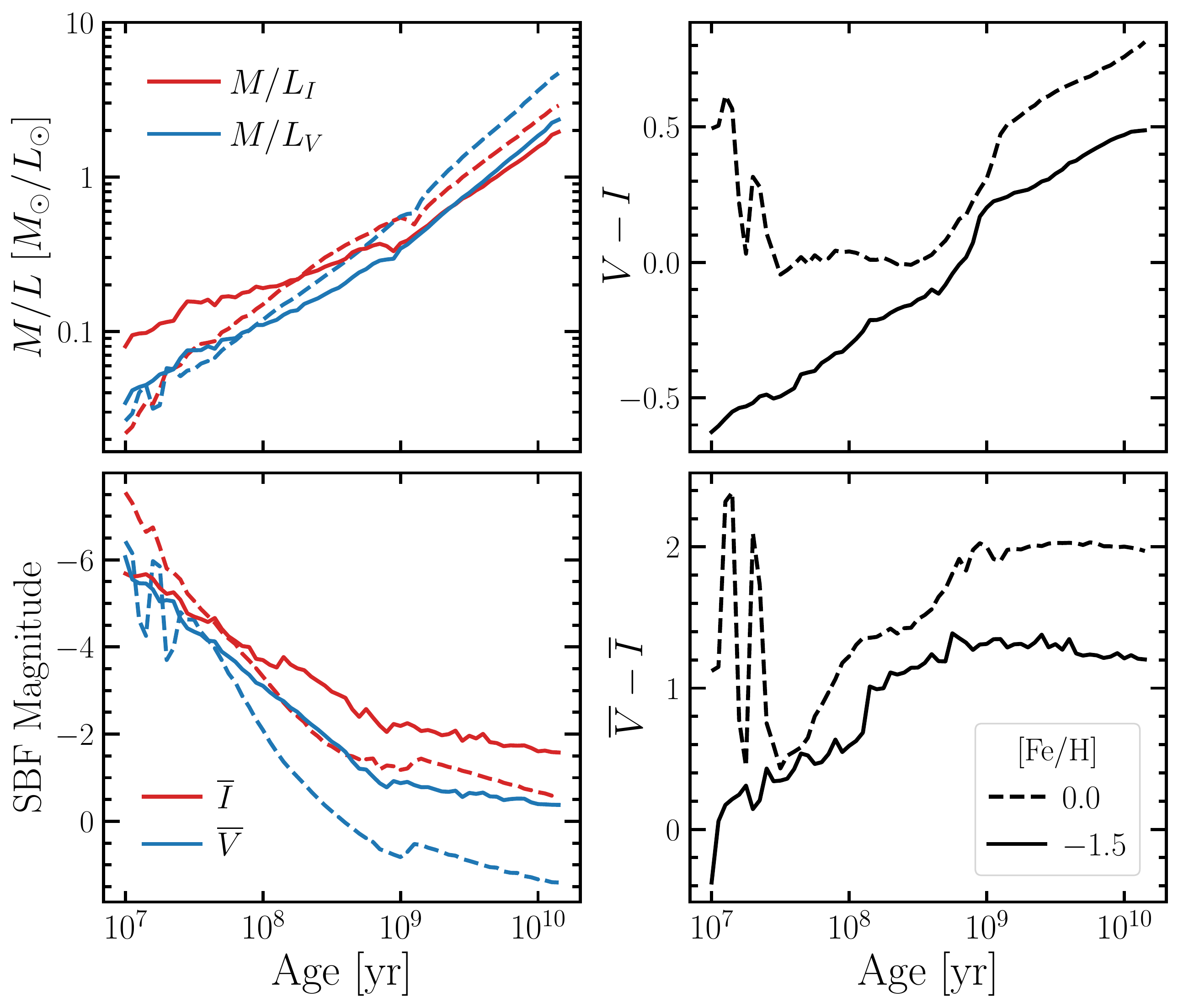}
  \caption{\code{ArtPop} calculations of the time evolution of SSPs. The top row shows the $V$- and $I$-band mass-to-light ratios (left) and integrated $V-I$ color (right). The bottom row shows the $V$- and $I$-band absolute SBF magnitudes (left) and $\overline{V} - \overline{I}$ SBF color (right). The SSPs were generated using the \citet{Kroupa:2001} initial mass function. Solar metallicity SSPs are shown as dashed lines, and metal poor SSPs with $[\mathrm{Fe/H}]=-1.5$ are shown as solid lines. \script{figure_2_greco_and_danieli.py}}
  \label{fig:sps}
\end{figure}

In Figure~\ref{fig:sps}, we show the results from \artpop calculations of the time evolution of an SSPs mass-to-light ratio (top left), $V-I$ color (top right), $I$- and $V$-band SBF magnitude, and $\overline{V}-\overline{I}$ SBF color. The calculations are performed using the \code{MISTIsochrone} class. As with all figures in this paper, a link to the code used to create the figure is provided in the caption. We note that we carried out a detailed comparison between the stellar population synthesis calculations of \artpop and the Flexible Stellar Population Synthesis software package \citep{Conroy:2009, Conroy-Gunn-2010} and found consistent results when care was taken to control for all model differences (e.g., spectral library, filter throughput functions, and mass definitions). 

The above calculations were performed using integrals over the full IMF. To build stellar populations composed of a finite numbers of stars, \artpop samples stellar masses from the IMF and interpolates stellar fluxes from a stellar isochrone, as described in Section~\ref{sec:pops}. This task is performed using the \code{SSP} class, which takes an \code{Isochrone} object, the IMF, and the number of stars (or total stellar mass) in the system:
\begin{lstlisting}[language=Python]
from artpop.stars import SSP

ssp = SSP(
    isochrone=iso,  # Isochrone object
    num_stars=1e5,  # number of stars
    imf="salpeter", # initial mass function
)
\end{lstlisting} 
Similar to the \code{MISTIsochrone} class, there is a \code{MISTSSP} helper class, which loads a MIST isochrone for a given set of SSP parameters. The above \code{ssp} object has various methods for calculating integrated properties. For example, the $i$-band magnitude and $g-i$ color are calculated using the \code{total\_mag} and \code{integrated\_color} methods: 
\begin{lstlisting}[language=Python]
i = ssp.total_mag("LSST_i")
g_i = ssp.integrated_color("LSST_g", "LSST_i")
\end{lstlisting}
The distance of the population is set to 10~pc by default, so the above magnitude is in absolute units. 

To build composite stellar populations (CSPs) in \artpop, \code{ssp} objects may be intuitively added together using the \code{+} operator. For example, suppose we have created two SSPs, one old (\code{ssp\_old}) and the other young (\code{ssp\_young}). Then, we may combine them into a single composite population as follows:
\begin{lstlisting}[language=Python] 
csp = ssp_old + ssp_young
\end{lstlisting}
The new \code{csp} object is a composite of the old and young SSPs, inheriting all the same methods for calculating integrated properties.

We emphasize that, since \code{SSP} objects contain a finite number of stars, the integrated properties will be stochastic due to incomplete sampling of the mass function \citep[e.g.,][]{Santos:1997, Greco:2021}. The number of stars required for the calculations to converge is a function of stellar population parameters (e.g., due to the frequency of rare, luminous stars) and photometric bandpass, but in general it takes a total stellar mass of ${>}10^6$~M$_\odot$ to approach a fully sampled mass function. However, if the goal is to fully sample the IMF, the \code{Isochrone} class should be used.

\begin{figure*}[t!]
  \centering
  \includegraphics[width=\textwidth]{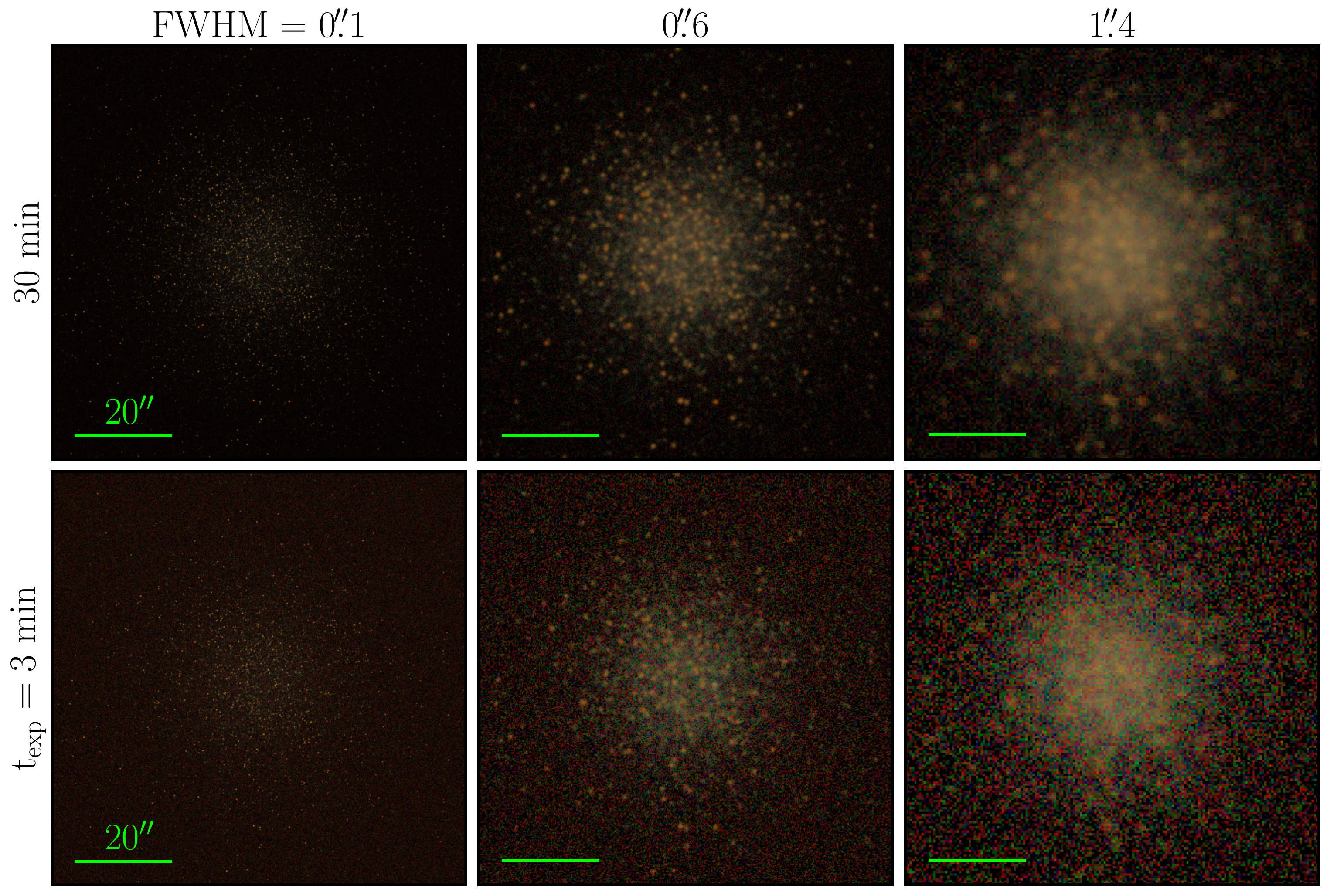}
  \caption{\artpop simulations of an old dwarf spheroidal of stellar mass $3\times10^6~\mathrm{M}_{\odot}$, placed at a distance of $5\,\mathrm{Mpc}$. The galaxy is composed of an SSP of age $12.6\,\mathrm{Gyr}$ and metallicity $\mathrm{[Fe/H]} = -1.8$. We show $gri$-composite images of mock observations of varying image resolution (FWHM values indicated at the top of each column) and exposure times (indicated on the left of each row). The image resolutions were chosen to be similar to HST/ACS (left column), HSC (middle column), and SDSS (right column). All images assume a mirror diameter of 8~m and sky brightnesses of 22, 21, and 20 mag~arcsec$^{-2}$ in $g$, $r$, and $i$, respectively. In each panel, the green line indicates the scale of 20\arcsec.  \script{figure_3_greco_and_danieli.py}}
  \label{fig:exptime_seeing}
\end{figure*}

\begin{figure*}[t!]
  \centering
  \includegraphics[width=\textwidth]{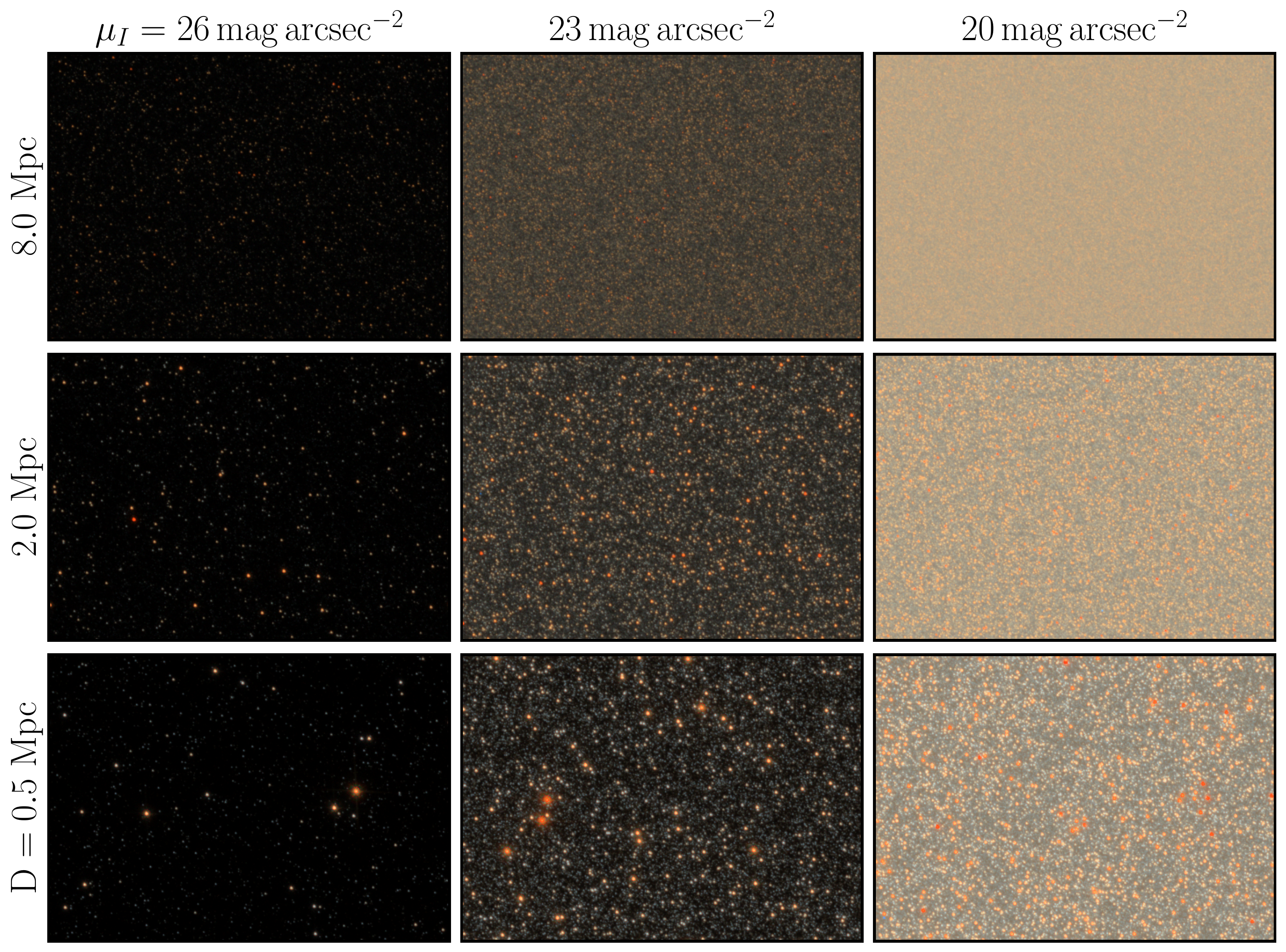}
  \caption{Simulated HST ACS/WFC $grI$-composite images of uniformly distributed SSPs with $\log(\mathrm{Age})=\mathrm{10}\,\mathrm{Gyr}$ and $[\mathrm{Fe}/\mathrm{H}]=-1.6$. The mean $I$-band surface brightness of each column is indicated at the top, and the distance to the populations in each row is indicated on the left. This figure visually demonstrates the well-known result that surface brightness is independent of distance (in the nearby universe). The images are $35\arcsec\times\,25\arcsec$, with a pixel scale of 0.05~arcsec~pixel$^{-1}$. \script{figure_4_greco_and_danieli.py}}
  \label{fig:sb}
\end{figure*}

\subsection{Coding Example: Image Simulations}\label{sec:code-image}

To create artificial images, \artpop uses \code{Imager} objects, which are implemented in the \code{artpop.image} module. There are two types of \code{Imager} objects: \code{IdealImager} and \code{ArtImager}. The former generates noiseless images, and the latter generates fully artificial images that include simulated noise from the sky, source, and detector. Initializing an \code{ArtImager} object therefore requires both instrumental (e.g., mirror diameter and read noise) and observational (e.g., exposure time and the sky surface brightness) parameters as input.

Similar to real observatories, a single \code{Imager} object is designed to ``observe'' any number of sources. This is convenient when the goal is to mock observe many sources using the same imaging setup. In \artpop, sources are stored in \code{Source} objects, which are containers that hold the positions and magnitudes of the stars. A primary purpose of \artpop is to generate these positions and magnitudes, though this is not necessary to create images using \code{Imager} and \code{Source} objects. As a simple example, let us create a mock observation using the SSP created in Section~\ref{sec:code-ssp}. 

First, we create a \code{Source} object using the magnitudes from the previously created SSP. For this example, we use the \code{artpop.plummer\_xy} function to sample positions from a Plummer distribution \script{section_3_greco_and_danieli.py}:
\begin{lstlisting}[language=Python]
from astropy import units as u
from artpop import Source
from artpop.space import plummer_xy

# image dimensions
xy_dim = (501, 501)

pixel_scale = 0.2 * u.arcsec / u.pixel

# returns a 2D numpy array
xy = plummer_xy(
    num_stars=ssp.num_stars,
    scale_radius=500*u.pc,
    distance=8*u.Mpc,
    xy_dim=xy_dim,
    pixel_scale=pixel_scale
)

# ssp magnitudes stored in astropy table
src = Source(
    xy=xy,              # image coordinates
    mags=ssp.mag_table, # magnitude table
    xy_dim=xy_dim       # image dimensions
)\end{lstlisting}
The above \code{src} object holds the stellar positions and magnitudes for a system of 10$^5$ stars (the number of stars in \code{ssp}) at a distance of 8~Mpc, with a spatial distribution that follows a Plummer profile of scale radius 500~pc. We also specify the image dimensions and pixel scale in order to convert the positions to image coordinates and flag stars that fall within the image. Stars that fall outside the image contribute to the total mass but are masked within the array of stellar positions, since they will not be injected into the image.

For simplicity, we will observe the source using an \code{IdealImager}, which may be initialized without any input parameters:
\begin{lstlisting}[language=Python]
from artpop.image import IdealImager

imager = IdealImager()
\end{lstlisting} 
Mock observations are carried out using the \code{observe} method. Here, we will observe the artificial source in the $i$ band, assuming 0\farcs6 seeing, which we will model as a Moffat profile using the \code{moffat\_psf} function:
\begin{lstlisting}[language=Python]
from artpop.image import moffat_psf

# returns a 2D numpy array
psf = moffat_psf(
    fwhm=0.6*u.arcsec, 
    pixel_scale=pixel_scale
)

obs = imager.observe(src, "LSST_i", psf)
\end{lstlisting} 
The returned object, \code{obs}, is an \code{IdealObservation} object, which is a container that holds the PSF-convolved image, as well as metadata such as the zero point and observation bandpass.

The above examples show that multiple steps are required to create stellar positions and magnitudes, which are required to create a \code{Source} object. For convenience, \artpop provides helper classes for creating complete  \code{Source} objects in a single step. For example, a \code{Source} object composed of an SSP with a \sersic spatial distribution and synthetic photometry generated using the MIST isochrones can be created using the \code{MISTSersicSSP} class.

\section{Example Applications} \label{sec:examples}

There is a wide range of use cases for \artpop. From visualizing the age-metallicity degeneracy to measuring survey detection efficiencies to generating synthetic data for machine learning algorithms, its potential applications span both scientific and pedagogical projects. In this section, we present example \artpop applications that highlight different features of the code. Each example has a corresponding figure with a link to the code used to generate it in the caption.

\subsection{Image resolution and exposure time}\label{sec:resolution}

The \code{ArtImager} class generates fully artificial images, adding noise from the detector, sky, and artificial source according to the user-provided instrumental and observational parameters. For a fixed artificial source, this is particularly useful for visually and quantitatively exploring the interplay between observational parameters such as exposure time and image resolution.

In Figure \ref{fig:exptime_seeing}, we show $gri$-composite images of the same dwarf galaxy placed at 5~Mpc, with the pixel scales and resolutions similar to HST/ACS (left column), HSC (middle column), and SDSS (right column). The dwarf galaxy has a Plummer mass distribution with a scale radius of 400~pc and a total stellar mass of $3\times10^6$~M$_\odot$. The bottom row shows mock observations with an exposure time of 3~min, and the top row shows mock observations with a factor of 10 longer exposure time. Other than the resolutions and pixel scales, the mock observation setups are identical. 

Comparing the top and bottom rows, the increase in exposure time leads to the expected increase in the signal-to-noise ratio. Stars in the outskirts of the galaxy disappear below the noise level in the short exposure panels. The comparisons become more interesting
when we also vary the image resolution (both seeing and pixel scale). At the highest resolution, the RGB is resolved in the longer 30~min exposure, but the small pixels (0.05~arcsec~pixel$^{-1}$) make the source appear as a diffuse object. As the resolution is decreased, stars blend into brighter point sources and the full galaxy becomes more easily detected as a single coherent object. 

\vspace{0.5cm}

\subsection{Surface brightness and distance}\label{sec:sb}

\begin{figure*}[t!]
  \centering
  \includegraphics[width=\textwidth]{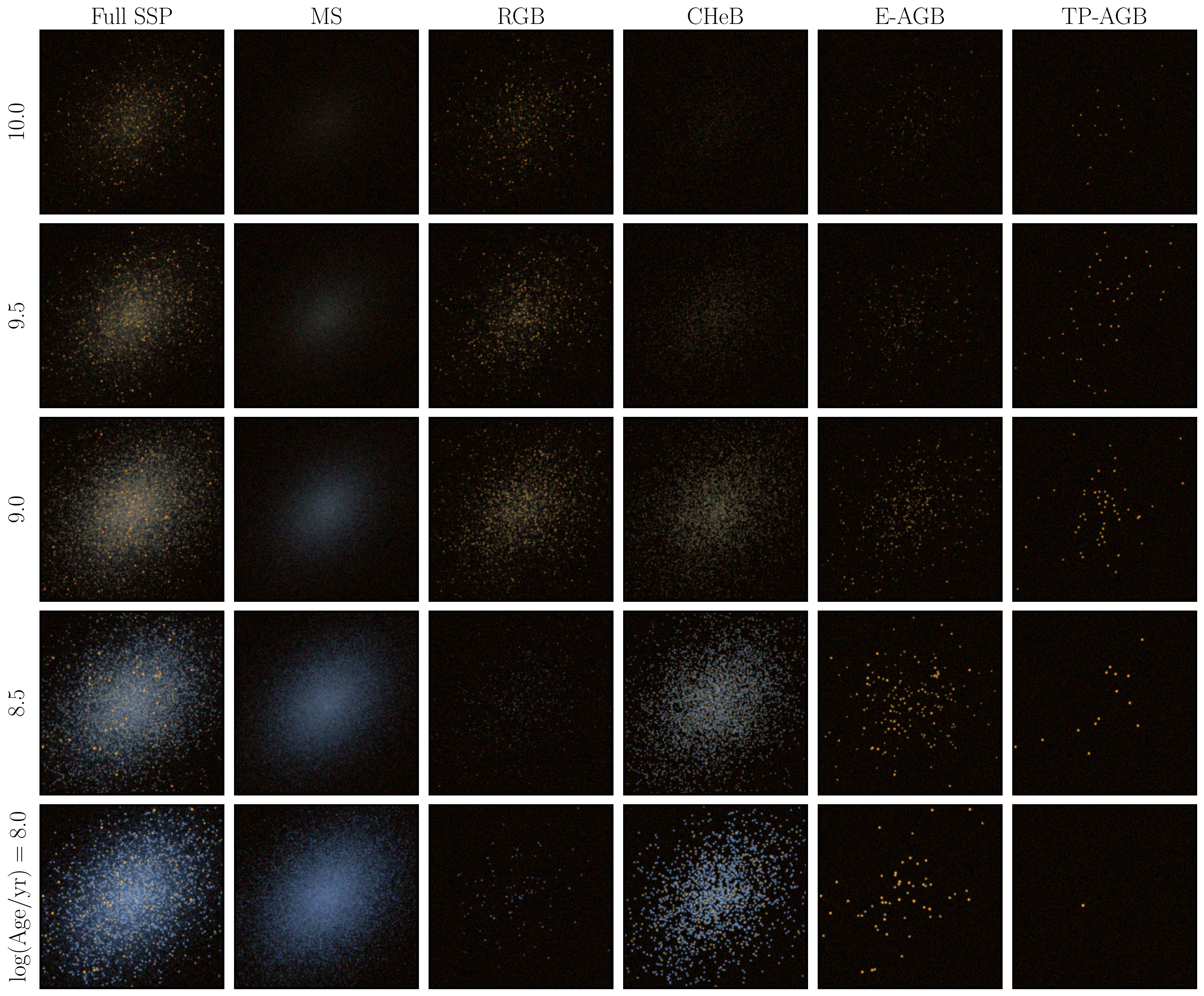}
  \caption{Simulated $gri$-composite images of dwarf galaxies of stellar mass 10$^6$~M$_\odot$ at a distance of 5 Mpc. The galaxies are SSPs with $\mathrm{[Fe/H] = -1.5}$, assuming the MIST model isochrones. Each row shows a single galaxy of fixed age, which is indicated on the left. For each galaxy, the leftmost panel shows the full SSP, and the remaining five panels show stars that are on the main sequence (MS), red giant branch (RGB), core-helium burning (CHeB) stars, and the early and thermally pulsating asymptotic branches (E-AGB and TP-AGB, respectively). The simulations were tuned to resemble an LSST-like observatory and observing conditions with exposure times of 90 minutes in $i$ and 45 minutes in $g$ and $r$. Each panel is 2~kpc on a side. As noted in the main text, the phases are defined according to the MIST primary equivalent evolutionary phases. \script{figure_5_greco_and_danieli.py}}
  \label{fig:phase-grid}
\end{figure*}

\begin{figure*}[t!]
  \centering
  \includegraphics[width=\textwidth]{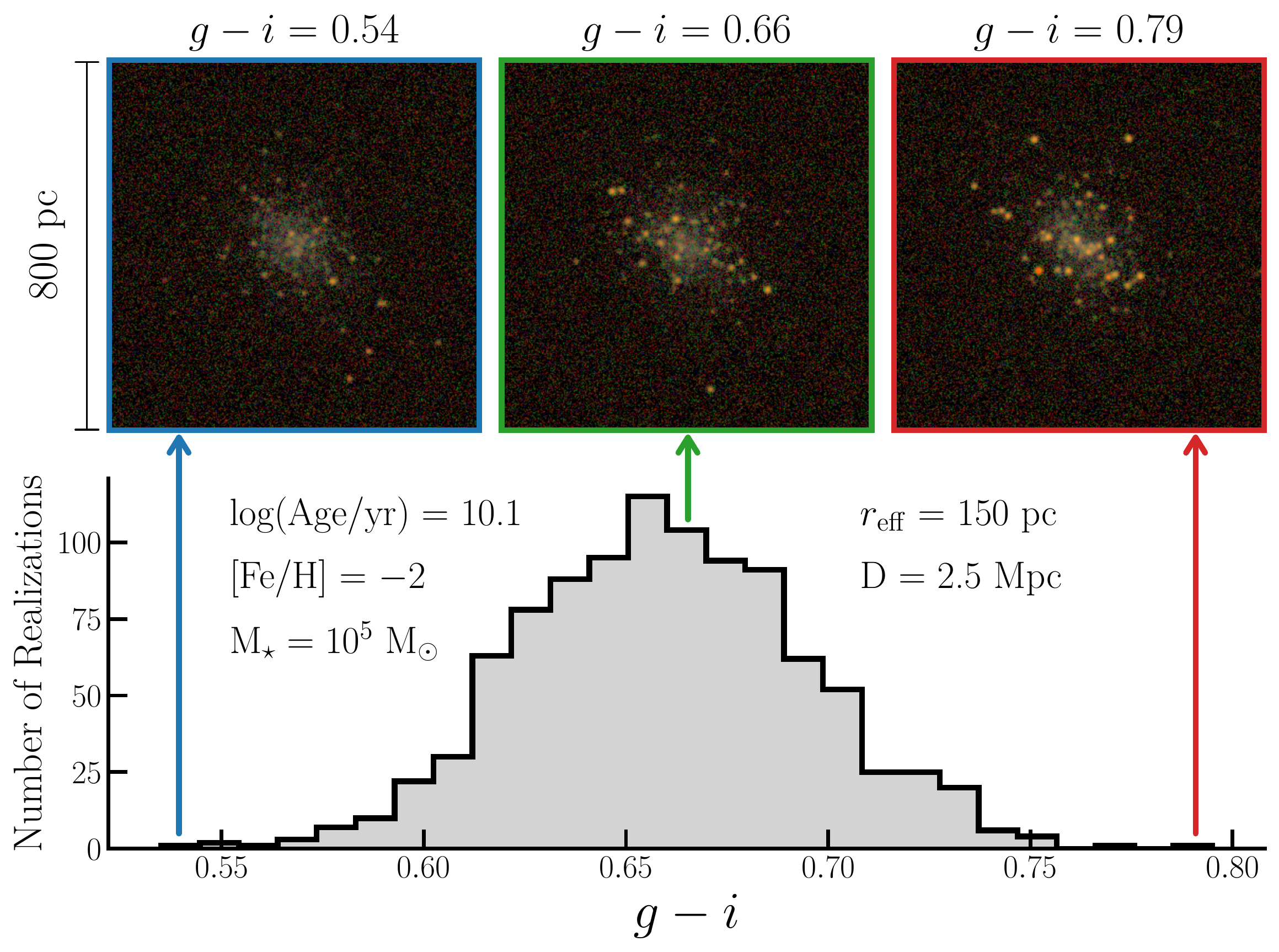}
  \caption{DECam $g-i$ color distribution of 1000 realizations of an \artpop dwarf galaxy with fixed model parameters. The galaxy is composed of an ancient, metal poor SSP with an age of 12.6~Gyr and metallicity of $\mathrm{[Fe/H] = -2}$. Given its low stellar mass of 10${^5}$~M$_\odot$, stochasticity in the numbers of luminous evolved stars leads to a wide range of integrated and visual properties. The $gri$-composite images on the top show the bluest (left), median (middle), and reddest (right) dwarf galaxy in the sample. A DECam-like observatory and observing conditions were assumed, with ${\sim}1\arcsec$ seeing and 2~hr (1~hr) exposures in $i$ ($g$ and $r$). The galaxy is placed at a distance of 2.5~Mpc, and the images are 800~pc on a side. \script{figure_6_greco_and_danieli.py}}
  \label{fig:dwarf-stoc}
  
\end{figure*}
\begin{figure*}[t!]
  \centering
  \includegraphics[width=\textwidth]{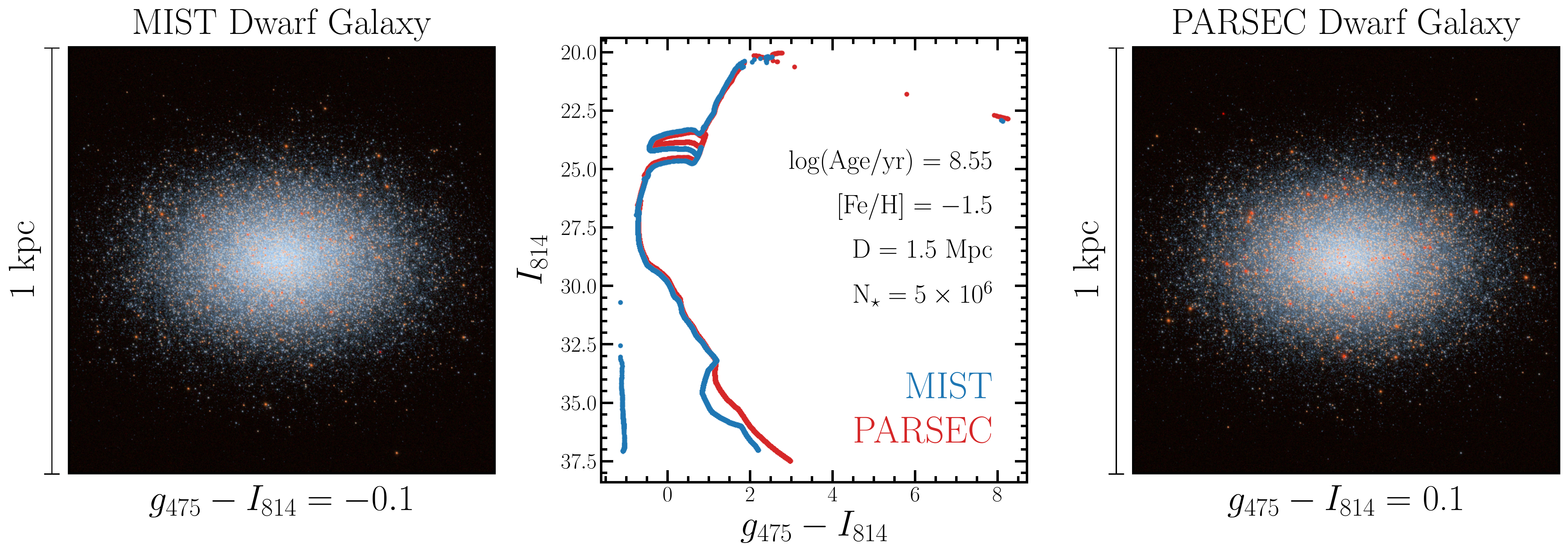}
  \caption{Visual comparison of artificial dwarf galaxies generated using stellar fluxes from the MIST (left panel) and PARSEC (right panel) isochrones. The isochrones were generated consistently using the Flexible Stellar Population Synthesis software package. Other than the choice of isochrone, the model parameters are identical, including the individual stellar positions and masses. The images are HST ACS/WFC $grI$-composites, with the integrated $g_{475}-I_{814}$ galaxy color indicated below each image. The middle panel shows the color-magnitude diagram of the constituent MIST (red) and PARSEC (blue) stars. The galaxies are SSPs composed of $5\times10^6$ stars with the population parameters indicated in the middle panel. \script{figure_7_greco_and_danieli.py}}
  \label{fig:mist-vs-parsec}
\end{figure*}

In the nearby universe, surface brightness is independent of distance. This important result can be visualized using \artpop's \code{MISTUniformSSP} class, which generates a uniformly distributed SSP using the MIST isochrone models and a user-specified average surface brightness and distance. Fixing the SSP parameters and field of view, the number of stars in an image increases as a function of distance, exactly compensating for dimming due to the inverse square law to ensure the surface brightness remains constant. 

In Figure~\ref{fig:sb}, we show image simulations of uniformly distributed SSPs with mean $I$-band surface brightnesses of 26 (left column), 23 (middle column), and 20 mag~arcsec$^{-2}$ (right column)---spanning the stellar density range from low-mass dwarf galaxies to globular clusters \citep[e.g.,][]{Munoz:2015}. We place the SSPs at distances of 0.5 (bottom row), 2 (middle row), and 8~Mpc (top row). In all panels, the stellar population age and metallicity  are 10~Gyr and [Fe/H]~$=-1.6$, respectively. Each image is $35\arcsec\times\,25\arcsec$, with a pixel scale of 0.05~arcsec~pixel$^{-1}$. For the mock observations, we used the \code{ArtImager} with 90~minute exposures in each of the HST ACS/WFC $I_\mathrm{814}$, $r_\mathrm{606}$, and $g_\mathrm{475}$ filters. The PSFs were generated using \code{Tiny Tim} \citep{Krist:1995}.

As expected, we see that fainter stars become increasingly resolved at fainter surface brightnesses (lower stellar density) and closer distances. At the nearest distance in the 23~mag~arcsec$^{-2}$ column (middle panel of the bottom row), there are ${\sim}3\times10^5$ stars, with the blue main sequence resolved into individual stars and only a handful of luminous giants in the frame. As the stars are placed to larger distances, their numbers increase as the square of the distance---there are ${\sim}5\times10^6$ stars in the 2~Mpc panel and ${\sim}8\times10^7$ in the 8~Mpc panel. At large distance and high surface brightness, Poisson fluctuations in the numbers of stars are too small to detect, leading to a visually smooth image. See \citet{Greco:2021} for a detailed study of these so-called surface brightness fluctuations using \artpop.

More than 200 million stars are required to generate the high surface brightness 20~mag~arcsec$^{-2}$ population at 8~Mpc, which is memory intensive. For such situations, \artpop provides an option so set a magnitude limit, fainter than which individual stars will not be sampled. Instead, the flux from the stars is combined into a smooth model, which is added to the image along with the brighter individual stars.

\subsection{Dwarf galaxies and stellar populations}\label{sec:stars}

\artpop makes it easy to visualize stellar systems as a function of astrophysical (e.g., distance, stellar mass, and SSP age) and observational (e.g., exposure time, bandpass, aperture diameter, and sky surface brightness) parameters. Moreover, since stars are injected individually, it is possible to visually compare how different phases of stellar evolution contribute to the (semi)resolved appearance and integrated properties of the system, which helps build intuition for interpreting images of similar systems in real data. 

To demonstrate the pedagogical utility of isolating stellar phases in artificial images, Figure~\ref{fig:phase-grid} shows $gri$-composite images of simulated dwarf galaxies at a distance of 5~Mpc. The distribution of stars in each galaxy follows a \sersic distribution, with a total stellar mass of 10$^6$~M$_\odot$. Each galaxy is composed of a metal-poor SSP with $\mathrm{[Fe/H] = -1.5}$ and an age ranging from 10~Gyr (top row) to 100~Myr (bottom row), assuming the MIST model isochrones. Each row shows the same galaxy realization, with the full SSP shown in leftmost panel. The remaining five panels show stars that are in the evolutionary phase indicated at the top of each column. From left to right, the phases are the main sequence (MS), red giant branch (RGB), core-helium burning (CHeB) stars, and the early and thermally pulsating asymptotic branches (E-AGB and TP-AGB, respectively). Similar to \citet{Greco:2021}, we label phases of stellar evolution according to the MIST primary equivalent evolutionary phases \citep[EEP;][]{Choi:2016, Dotter:2016}, which are useful computationally but in some cases lead to terminology that differs from standard nomenclature (e.g., an RGB phase associated with high-mass stars).

When the mass of a stellar system is $\lesssim10^5$~M$_\odot$ \citep{Greco:2021}, the IMF becomes significantly undersampled. At such low masses, the numbers of the most luminous stars range from zero to a dozen or so, leading to a wide range of integrated properties and visual appearance. Using \artpop, we can quantitatively and visually compare different mock galaxy realizations with identical stellar population and observational parameters. 

In Figure~\ref{fig:dwarf-stoc}, we show the DECam $g-i$ color distribution of 1000 realizations of an \artpop dwarf galaxy with fixed model parameters. The galaxy has an age of 12.6~Gyr, metallicity of $\mathrm{[Fe/H]=-2}$, and low stellar mass of 10$^5$~M$_\odot$. Stochasticity in the numbers of evolved stars---particularly AGB stars---results in a standard deviation of ${\sim}0.05$~mag and range of ${\sim}0.25$~mag in $g-i$ color. In the top three panels, we show $gri$-composite images of the bluest (left), median color (middle), and reddest (right) dwarf galaxy realization in the sample.

\subsection{Comparing isochrone models}\label{sec:iso_compare}

 With new imaging surveys like LSST and ultimately using the Roman Space Telescope, there will soon be a vast increase in the number of low-mass galaxies with high-quality imaging and (semi-)resolved stellar populations. These systems will span a large range of stellar population parameters, potentially providing powerful benchmarks for stellar population synthesis models. While model comparisons using color-magnitude diagrams are common, \artpop's modular design makes it straightforward to additionally generate artificial galaxy images based on the different models. 

In Figure~\ref{fig:mist-vs-parsec}, we show a visual comparison of dwarf galaxies generated using stellar fluxes from the MIST (left panel) and PARSEC \citep[right panel;][]{Bressan:2012} isochrone models. The shown HST ACS/WFC $grI$-composite images are based on mock observations with an HST-like observatory using 180~min exposures. To ensure the stellar fluxes were calculated consistently, we used the Flexible Stellar Population Synthesis software package \citep{Conroy:2009, Conroy-Gunn-2010}. The  center panel shows the MIST (blue points) and PARSEC (red points) color-magnitude diagrams associated with the stars in the mock images. 

Other than the isochrone model, all model parameters are identical---including the stellar positions and masses. The galaxies are composed of SSPs with $5\times10^6$ stars of age ${\sim}$355~Myr and metallicity $\mathrm{[Fe/H] = -1.5}$. To exactly match the stellar masses, we restricted the mass range from 0.1~M$_\odot$ to 2.8305~M$_\odot$. The minimum stellar mass is set by the MIST isochrones, whereas the maximum stellar mass is set by the PARSEC isochrones. 

\subsection{Injecting into real images}

\begin{figure*}[t!]
 \centering
 \includegraphics[width=\textwidth]{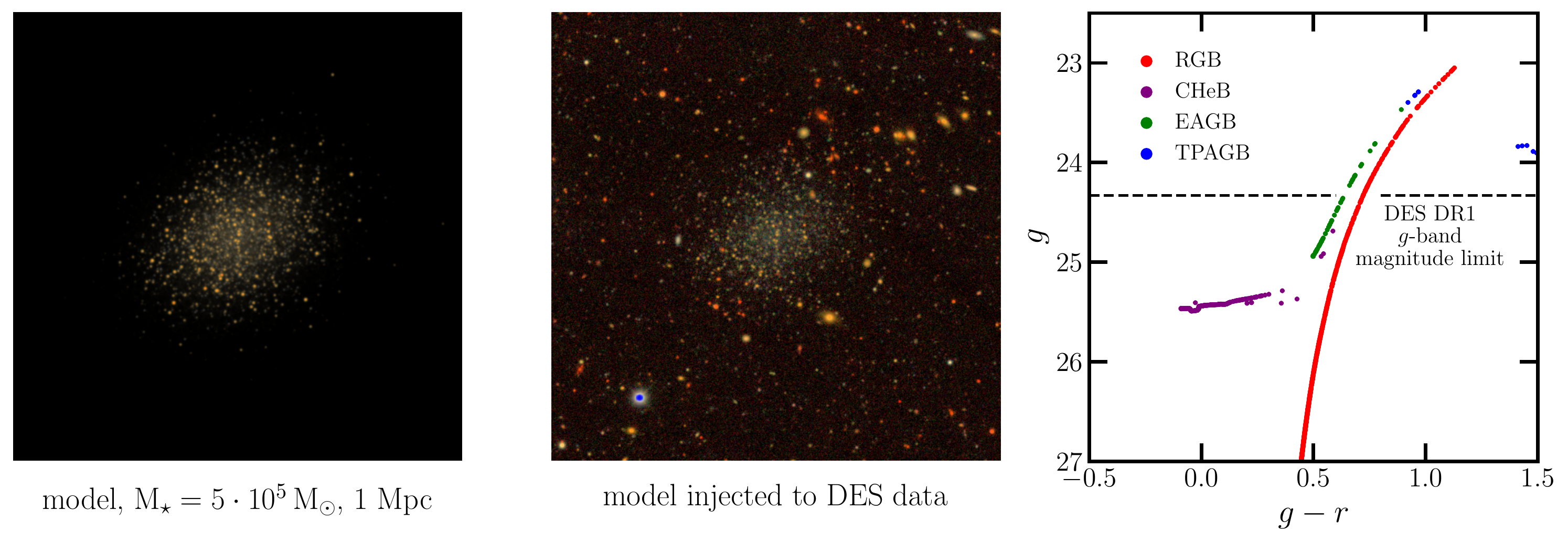}
  \caption{\artpop model of an ultra-faint dwarf galaxy with a stellar mass of $\mathrm{M}_{\star}=5 \cdot 10^5\,\mathrm{M}_{\odot}$, placed at a distance of $1\,\mathrm{Mpc}$. The left panel shows the gri-composite color image and the middle panel shows the same model injected into a Dark Energy Survey image. In the right panel we show the $gr$ color-magnitude diagram, where stars are color-coded according to their stellar phase. The dashed horizontal line marks the $g$-band limiting magnitude reported by DES DR1 of $m_{g,\mathrm{lim}}=24.33\,\mathrm{mag}$. Ignoring star-galaxy separation issues, stars that are brighter than this limit should, in theory, be detected in the image. \script{figure_8_greco_and_danieli.py}}
  \label{fig:dwarf_des}
\end{figure*}

A complementary approach to using the \code{ArtImager} class for generating fully artificial images (as shown in Section \ref{sec:resolution}-\ref{sec:iso_compare}) is to inject artificial sources into \textit{real} astronomical images. Using this approach,  \artpop has been proven an effective tool in estimating imaging survey depth and detection completeness (\citealt{Greene:2021}). Particularly, the dual-mode functionality of \artpop, namely generating stellar population models and artificial images, can be used simultaneously in the following way.   

Using the \code{MISTSersicSSP} class, we generate an artificial dwarf galaxy source, placed at $1\,\mathrm{Mpc}$, assuming an old SSP with $M_\star=5\cdot 10^5\,\mathrm{M}_\odot$, $\log (\mathrm{Age})=10.1\,\mathrm{Gyr}$, $\mathrm{[Fe/H]=-2}$, and a \sersic surface brightness distribution. We then use the \code{IdealImager} to generate a noiseless image of the source. Assuming DES-like observational parameters, we ``observe'' the galaxy in the $g$, $r$, and $i$ bands. The $gri$-composite noiseless image of the model is shown in the left panel of Figure \ref{fig:dwarf_des}. 

Next, we inject the noiseless model into a DES image\footnote{DES DR1 coadd tiles ($0.7306\,\mathrm{deg}$ on a side) were downloaded from the following public data server: \url{http://desdr-server.ncsa.illinois.edu/despublic/dr1_tiles/.}}. This is done by simply adding the image shown on the left panel to the a DES tile. The injected image is shown in the middle panel of Figure~\ref{fig:dwarf_des}. As expected, many of the stars that are visible in the outskirts of the noiseless model image blend into the noise in the middle image, though a small number of giants are visually detected. Finally, in the right panel, we show the CMD of the source using synthetic photometry in the DECam photometric system. The dashed black line marks the $g$-band limiting magnitude of the DES DR1. 

For understanding the detection of low stellar density systems in imaging surveys, models of dwarf galaxies, spanning a range of stellar masses, chemical compositions, ages, and morphologies can be generated. As demonstrated, \artpop provides both catalogs of stars that can be injected into existing survey catalogs, accounting for noise and detection limits, and realistic images with photometry in the appropriate photometric system that can be then injected into survey images.

\section{Summary} \label{sec:summary}

In this paper we have presented \artpop, a public software package for synthesizing of stellar populations and simulating realistic images of stellar systems. The code is modular and designed to allow maximal user flexibility. \artpop is under active development and currently provides the following capabilities:
\begin{enumerate}
    \item Stellar population synthesis: The \code{artpop.stars} module  builds simple and composite stellar populations by sampling a user-specified initial mass function. Stellar fluxes are calculated by interpolating pre-calculated magnitude grids from a stellar isochrone model, which the user is free to choose. \github{stars} \readthedocs{pops} 
    \item Sampling spatial distributions: The \code{artpop.space} module samples two-dimensional positions in image coordinates. Currently, we have implemented samplers for uniform, Plummer, and \sersic distributions. Grid sampling for arbitrary two-dimensional functions is also possible using \code{Astropy} model objects\footnote{\url{https://docs.astropy.org/en/stable/modeling/index.html}}. \github{space} \readthedocs{spatial} 
    \item Image simulations: The \code{artpop.image} module generates artificial images of \code{Source} objects. The simulations can be fully artificial images with realistic noise or ideal noiseless images, which may be injected into real imaging data. \github{image} \readthedocs{artimages} \readthedocs{inject} 
\end{enumerate}
These three functionalities can be used independent or collectively to generate catalogs and images of different stellar systems such as galaxies, globular clusters, and stellar streams. 

We encourage the reader to install \artpop, go through the tutorials on the website, and run some of the examples given in this paper. Installation instructions are given at \url{https://artpop.readthedocs.io/en/latest/getting_started/install.html}. Please feel free to report bugs and request features using the \artpop GitHub issues page or by submitting a pull request to make a code contribution. More information about contributing to \artpop can be found at \url{https://artpop.readthedocs.io/en/latest/getting_started/contribute.html}.

\software{
  \code{astropy} \citep{Astropy-Collaboration:2013aa}, 
  \code{Flexible Stellar Population Synthesis} \citep{Conroy:2009, Conroy-Gunn-2010},
  \code{matplotlib} \citep{Hunter:2007aa}, 
  \code{numpy} \citep{Van-der-Walt:2011aa}, 
  \code{scipy} (\url{https://www.scipy.org})
}

\acknowledgments
We are deeply indebted to the open-source astronomy and scientific Python communities. Indeed, we learned almost everything we know about developing open-source software from poking around public GitHub repositories. We thank Adrian Price-Whelan and Pieter van Dokkum for providing useful comments, and we thank Ava Polzin for her help generating HST PSFs. 

J.P.G. was supported by an NSF Astronomy and Astrophysics Postdoctoral Fellowship under award AST-1801921. S.D. is supported by NASA through Hubble Fellowship grant HST-HF2-51454.001-A awarded by the Space Telescope Science Institute, which is operated by the Association of Universities for Research in Astronomy, Incorporated, under NASA contract NAS5-26555.

\bibliography{artpop_bib}
\bibliographystyle{aasjournal}

\end{document}